\documentclass[aps,pra,10pt,twocolumn,notitlepage,floatfix,superscriptaddress]{revtex4-1} 

\usepackage[dvips]{graphicx}
\usepackage{subfigure}
\usepackage{amsmath}
\usepackage{amssymb}
\usepackage{bm}
\usepackage{color}
\usepackage[colorlinks=true,citecolor=blue,urlcolor=blue,linkcolor=blue,hyperfigures=true]{hyperref}
\usepackage{multirow}
\usepackage[colorinlistoftodos]{todonotes}
\usepackage{comment}

\graphicspath{{./}}

\def\Fig{Fig.~}

\def\Ref{Ref.~}

\def\Sec{Sec.~}

\def\be{\begin{equation}}
\def\ee{\end{equation}}
\def\bea{\begin{eqnarray}}
\def\eea{\end{eqnarray}}

\def\eg{\textit{e.g.}~}

\def\bkeff{\bm{k}_{\rm eff}}

\newcommand{\ket}[1]{\left| #1 \right\rangle}

\begin{document}

\title{Generation of high-purity, low-temperature samples of $^{39}$K\\for applications in metrology}

\author{L. Antoni-Micollier}
\affiliation{LP2N, IOGS, CNRS and Universit\'{e} de Bordeaux, rue Fran\c{c}ois Mitterrand, 33400 Talence, France}
\author{B. Barrett}
\affiliation{LP2N, IOGS, CNRS and Universit\'{e} de Bordeaux, rue Fran\c{c}ois Mitterrand, 33400 Talence, France}
\affiliation{iXblue, 34 rue de la Croix de Fer, 78105 Saint-Germain-en-Laye, France}
\author{L. Chichet}
\affiliation{LP2N, IOGS, CNRS and Universit\'{e} de Bordeaux, rue Fran\c{c}ois Mitterrand, 33400 Talence, France}
\author{G. Condon}
\affiliation{LP2N, IOGS, CNRS and Universit\'{e} de Bordeaux, rue Fran\c{c}ois Mitterrand, 33400 Talence, France}
\author{B. Battelier}
\affiliation{LP2N, IOGS, CNRS and Universit\'{e} de Bordeaux, rue Fran\c{c}ois Mitterrand, 33400 Talence, France}
\author{A. Landragin}
\affiliation{LNE-SYRTE, l'Observatoire de Paris, PSL Research University, CNRS, Sorbonne Universit\'{e}s,\\UPMC Univ.~Paris 06, 61 avenue de l’Observatoire, 75014 Paris, France}
\author{P. Bouyer}
\affiliation{LP2N, IOGS, CNRS and Universit\'{e} de Bordeaux, rue Fran\c{c}ois Mitterrand, 33400 Talence, France}

\date{\today}

\begin{abstract}
We present an all optical technique to prepare a sample of $^{39}$K in a magnetically-insensitive state with 95\% purity while maintaining a temperature of 6 $\mu$K. This versatile preparation scheme is particularly well suited to performing matter-wave interferometry with species exhibiting closely-separated hyperfine levels, such as the isotopes of lithium and potassium, and opens new possibilities for metrology with these atoms. We demonstrate the feasibility of such measurements by realizing an atomic gravimeter and a Ramsey-type spectrometer, both of which exhibit a state-of-the-art sensitivity for cold potassium.
\end{abstract}

\maketitle

\section{Introduction}
\label{sec:Introduction}

Ultra-cold atomic sources with high state purity are the basis of many experiments that aim to study fundamental physics through the use of quantum simulators, or via precision measurements using atom interferometers \cite{Borde1989, Kasevich1991, Gillot2014, Rosi2014, Kovachy2015, Dutta2016} or atomic clocks \cite{Kasevich1989, Gibble1993, Santarelli1999, Nicholson2012, Hinkley2013}. These fields have grown more interested in exotic species that exhibit interesting properties that can be exploited to study new physical phenomena. However, lighter atoms such as the isotopes of lithium and potassium are rarely proposed as metrological tools. This is primarily because these atoms have closely-spaced energy structures and relatively large recoil velocities---making them notoriously difficult to cool and manipulate \cite{Cataliotti1998, Fort1998, Modugno1999, Landini2011, Gokhroo2011, Nath2013}. Additionally, their compact hyperfine splitting and lighter mass makes them much more sensitive to parasitic effects, such as AC Stark shifts, Zeeman shifts and forces due to electro-magnetic fields. Nevertheless, in some cases these properties can be advantageous---for instance, a larger sensitivity to atomic recoil could lead to a new determination of the fine structure constant \cite{Bouchendira2011, Cassella2016}. Similarly, precision measurements of different hyperfine splittings can shed new light on collisional interactions \cite{Gibble1995, Kokkelmans1997} and the variation of fundamental constants \cite{Blatt2008, Rosenband2008}.

The starting point for many experiments aimed at studying fundamental physics is to prepare a pure sample in terms of its energy, spin and momentum before injecting into an atom interferometer, spectrometer or quantum simulator. Previously, ultra-cold potassium isotopes have been spin-polarized via optical pumping \cite{Modugno2001, Roati2002, Roati2007, Fattori2008, Campbell2010}, adiabatic radio-frequency sweeps \cite{Wille2008}, and spin relaxation in a two-component mixture \cite{Spiegelhalder2010}. However, the majority of preparation schemes demonstrated so far have been carried out on magnetically-sensitive states which are placed in optical or magnetic traps for evaporation to quantum degeneracy. For applications in metrology, it is usually more desirable to use a magnetically-insensitive state to carry out measurements, yet this is a challenging task for species like potassium.

In this work, we report on an all-optical technique to prepare $^{39}$K atoms in the $\ket{F = 1, m_F = 0}$ state with $\sim 95$\% purity, while maintaining a temperature of 6 $\mu$K from a sample initially cooled to $\sim 4$ $\mu$K in a gray molasses. We use this high-purity, low-temperature atomic source in an atom-interferometric gravimeter \cite{LeGouet2008}, where we achieve a sensitivity to gravitational acceleration of 60 $\mu$Gal (1 $\mu$Gal = $10^{-8}$ m/s$^2$) after 4800 s of integration---representing the state-of-the-art achieved with potassium so far. We also perform Ramsey-type hyperfine spectroscopy \cite{Ramsey1949, Kasevich1989} with this source and demonstrate a sensitivity of $4.1 \times 10^{-11}$ to the ground state splitting of $^{39}$K, which corresponds to more than a factor of 30 improvement over previous measurements \cite{Arimondo1977}.

The rest of this article is organized as follows. In \Sec \ref{sec:Sequence}, we overview the laser-cooling and purification process. In \Sec \ref{sec:Metrology} we discuss experimental results pertaining to two ongoing precision measurements with $^{39}$K. Finally, we give our conclusions and perspectives in \Sec \ref{sec:Conclusion}.

\section{Cooling and purification}
\label{sec:Sequence}

\begin{figure*}[!tb]
  \centering
  \includegraphics[width=0.98\textwidth]{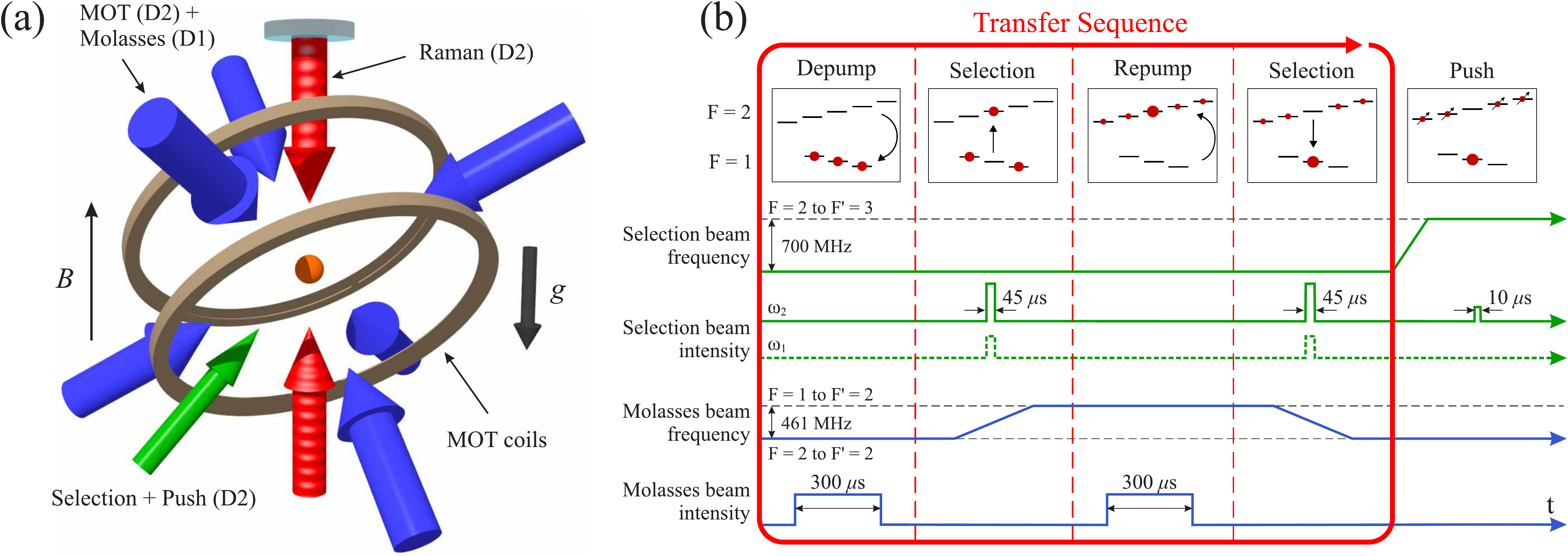}
  \caption{(a) Orientation of optical beams for cooling, state preparation and interrogation of the atoms. Light from both D1 and D2 lasers is overlapped with the same polarization along six beams used for the MOT loading, molasses cooling and optical pumping stages. The selection beam (D2 light) is configured as either a push beam to remove atoms in $\ket{F = 2}$, or as a co-propagating Raman beam to make coherent two-photon transitions with circular polarization. The counter-propagating Raman beams (D2 light) have perpendicular linear polarizations in order to enhance velocity-sensitive transitions and suppress velocity-insensitive ones. These beams are also used to detect the internal state of the atoms via resonant fluorescence on the $\ket{F = 2} \to \ket{F' = 3}$ transition. An external pair of Helmholtz coils (not shown) are used to generate a uniform magnetic bias field $\bm{B}$ along the vertical direction during the selection and interrogation phases. (b) State preparation sequence. The function of each optical pulse on the atomic states is shown in the first row. The relative intensities and frequencies of the selection and molasses beams are shown in the remaining rows. Note that the selection beam has two frequencies (labelled as $\omega_1$ and $\omega_2$) to make Raman transitions. The steps circled in the red loop indicate the transfer sequence, which is repeated to increase the fraction of atoms in the target state. At the end of this loop, a near-resonant push beam removes atoms remaining in $\ket{F = 2}$ to improve the purity of the target state.}
  \label{fig:KBeams+StatePrepSequence}
\end{figure*}

The first magneto-optical traps (MOTs) with the isotopes of potassium were achieved in the late 1990s \cite{Williamson1995, Wang1996, Cataliotti1998, Modugno1999}. Cooling and manipulation of potassium is challenging primarily as a result of the small energy separation between hyperfine levels in the $4^2 P_{3/2}$ excited state. This effectively opens the ``closed'' transitions on the D2 line that are largely forbidden in other alkali atoms, such as rubidium and cesium, and creates a complex sub-Doppler cooling mechanism that involves all of the hyperfine excited states. Various sub-Doppler cooling schemes have previously been explored using standard red-molasses techniques \cite{Fort1998, Landini2011, Gokhroo2011} that have achieved $^{39}$K temperatures as low as 25 $\mu$K. Recently, gray-molasses cooling \cite{Boiron1995} has been demonstrated as an efficient mechanism for alkali atoms with compact level structures, such as the isotopes of lithium \cite{Grier2013, Burchianti2014} and potassium \cite{RioFernandes2012, Nath2013, Salomon2013, Sievers2015}. This sub-Doppler cooling method, which combines the Sisyphus effect \cite{Dalibard1989} with velocity-selective coherent population trapping \cite{Aspect1989}, requires a light source tuned to the blue of the D1 transition. So far, temperatures as low as 6 $\mu$K have been demonstrated in $^{39}$K samples using this technique \cite{Salomon2013}. More recently, an ultra-low temperature of 1.8 $\mu$K was achieved using degenerate Raman sideband cooling in an optical lattice \cite{Grobner2017}. These conditions are ideal for further cooling large fractions to quantum degeneracy, particularly with all-optical techniques \cite{Salomon2014}.

In this work, we have adopted a cooling scheme similar to that of \Ref \cite{Salomon2013}. Both our D1 (770 nm) and D2 (767 nm) light sources are coupled through the same fiber splitter to generate a 6-beam MOT in the 1-1-1 configuration, as shown in \Fig \ref{fig:KBeams+StatePrepSequence}(a). Potassium-39 atoms are loaded in a 3D MOT from a background vapor in 0.5 s. We then switch off the magnetic gradient coils and the D2 light, and switch on the D1 light to cool the atoms in a gray molasses for 7 ms. During this time, the D1 cooling beam is blue detuned by 20 MHz from the $\ket{F=2} \to \ket{F'=2}$ transition, while the repump beam is detuned by the same amount from the $\ket{F=1} \to \ket{F'=2}$ transition---satisfying the critical Raman resonance condition for efficient gray-molasses cooling. To reduce additional photon scattering during the molasses, the cooling and repump beam intensities are decreased linearly from $18.5 \, I_{\rm{sat}}$ to $2.4 \, I_{\rm{sat}}$ and from $4.5 \, I_{\rm{sat}}$ to $0.6 \, I_{\rm{sat}}$, respectively ($I_{\rm sat} = 1.75$ mW/cm$^2$ \cite{Tiecke2011}). At the end of the molasses, we apply a short ($\sim 300$ $\mu$s) pulse using the D1 cooling light tuned near the $\ket{F=2} \to \ket{F'=2}$ transition in order to depump the atoms into $\ket{F = 1}$. We obtain a sample of $\sim 6 \times 10^7$ atoms at a temperature of $4.3 \pm 0.2$ $\mu$K, roughly equally distributed in population among the magnetic sub-levels of $\ket{F = 1}$. We verify the temperature of the sample using two methods; (i) by using time-of-flight (TOF) images recorded with a charge-coupled device (CCD) camera, and (ii) using velocity-sensitive two-photon Raman spectroscopy \cite{Moler1992} along the vertical direction \footnote{The Raman spectroscopy method of measuring temperature requires an increase in the bias field to $\sim 1$ G in order to resolve the velocity distribution of each magnetic sub-level.}.

\begin{figure*}[!tb]
  \centering
  \includegraphics[width=0.98\textwidth]{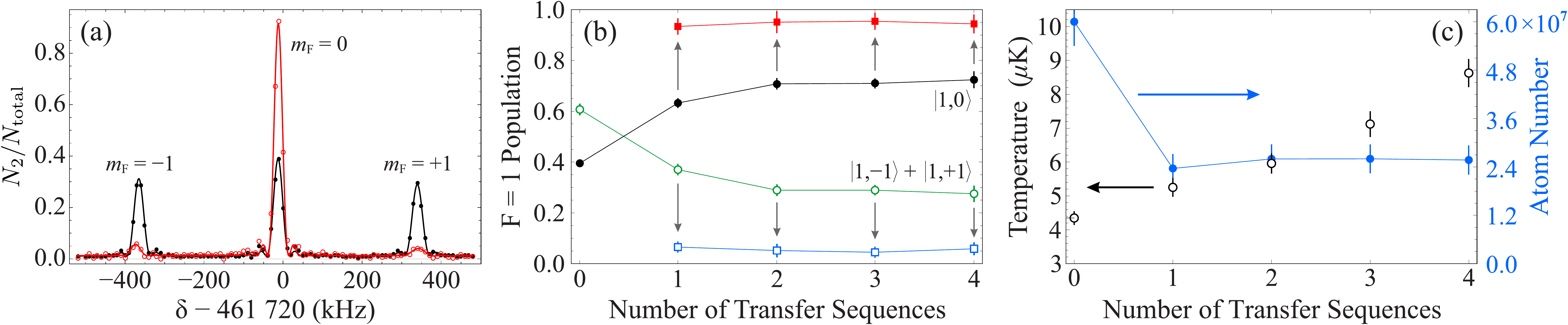}
  \caption{(a) Velocity-insensitive Raman spectrum of the sample before (black filled circles) and after (red open circles) state preparation using a single transfer sequence and a purifying push pulse. These data were obtained with a circularly-polarized Raman beam and by blocking the retro-reflected light along the vertical axis [see \Fig \ref{fig:KBeams+StatePrepSequence}(a)]. We estimate the relative population in each state from the amplitude obtained by least-squares fits to each resonance using a sinc$^2(\delta)$ lineshape. (b) Population in the target state $\ket{1,0}$ (filled symbols) and the total population in $\ket{1,\pm 1}$ (open symbols) as a function of the number of transfer sequences. Circles correspond to measurements made after $N$ transfer sequences with the push pulse replaced by a depump pulse [see \Fig \ref{fig:KBeams+StatePrepSequence}(b)]. Square symbols represent separate measurements including the push pulse, which acts to further purify the sample in the target state. (c) Temperature (open circles) and atom number (filled circles) versus the transfer sequence number. The initial decrease in atom number is due to the push pulse, which is applied only for $N > 0$, but thereafter remains roughly constant. The linear increase in temperature is due to the additional D1 optical pumping pulses made with each transfer sequence.}
  \label{fig:KSampleProperties}
\end{figure*}

Although a sub-Doppler-cooled sample is sufficient to proceed with some types of measurements \cite{Barrett2011a}, for a broad variety of experiments a distributed population amongst magnetic sub-levels is undesirable and a purification step is often employed. However, manipulating the state populations of potassium (along with other species exhibiting a light mass and a quasi-open set of electric dipole transitions) is complicated by its high susceptibility to scattering-induced heating and optical pumping. Additionally, the use of microwave transitions can be frustrated by the size and material of the vacuum system. For instance, the $\ket{F = 1, m_F = 0} \to \ket{F = 2, m_F = 0}$ clock transition in $^{39}$K at 461.7 MHz has a relatively large wavelength of $\sim 65$ cm, and small-volume metal chambers cannot support this field. We circumvent these challenges by combining optical Raman pulses to select atoms in the desired $m_F = 0$ state with optical pumping between hyperfine manifolds using near-resonant D1 light. The optical selection pulse plays the role of a microwave, since one can induce coherent state inversion with high efficiency and the optical fields can be kept far from single-photon resonances so as to avoid spontaneous emission. The advantage of optical pumping on the D1 line is that there are no cycling transitions, hence atoms scatter a minimum number of photons before falling into the desired ground state manifold.

Our state preparation sequence is outlined in \Fig \ref{fig:KBeams+StatePrepSequence}(b). We first apply a magnetic bias field of 0.25 G along the vertical axes in order to quantize and separate the magnetic sub-levels by a few hundred kHz. This field is servo-locked by measuring the field near the atoms with a flux-gate sensor (Bartington MAG-03MCTPB500) and feeding back on the current driving a large pair of compensation coils around the chamber. The sample is initially depumped to $\ket{F = 1}$ using D1 light along the MOT beams. Atoms in the $\ket{F = 1, m_F = 0}$ state are then transferred coherently to $\ket{F = 2, m_F = 0}$ using a co-propagating Raman $\pi$-pulse \footnote{To make $\Delta m_F = 0$ transitions, the two co-propagating Raman beams used for selection must have the same circular polarization, with a component of the electric field perpendicular to the quantization axis defined by the bias magnetic field.}. The atoms remaining in the $\ket{F=1}$ manifold are then optically pumped to $\ket{F = 2}$ using a repump pulse resonant with the $\ket{F = 1} \to \ket{F' = 2}$ D1 transition. Since the optical pumping light is isotropically polarized, this operation distributes the population remaining in $\ket{F = 1}$ roughly equally among the Zeeman states in $\ket{F = 2}$. We then drive the $\ket{F = 2, m_F = 0} \to \ket{F = 1, m_F = 0}$ transition with a second Raman pulse. This ensures that the accumulation of atoms in $m_F = 0$ from the optical pumping pulses is not reversed by subsequent pulses. Henceforth, we refer to this four-step process as a ``transfer'' sequence, as depicted in \Fig \ref{fig:KBeams+StatePrepSequence}(b). This transfer sequence can be repeated a number of times to further increase the fraction of atoms in the target state $\ket{F = 1, m_F = 0}$, but at the expense of additional heating to the sample. Finally, we apply a short pulse of D2 light resonant with the $\ket{F = 2} \to \ket{F' = 3}$ transition to push away any atoms remaining in $\ket{F = 2}$. We emphasize that this technique can easily be adapted to transfer atoms into any other target state with a suitable choice of the selection beam frequency, polarization and orientation relative to the $B$-field axis.

\begin{figure*}[!tb]
  \centering
  \includegraphics[width=0.98\textwidth]{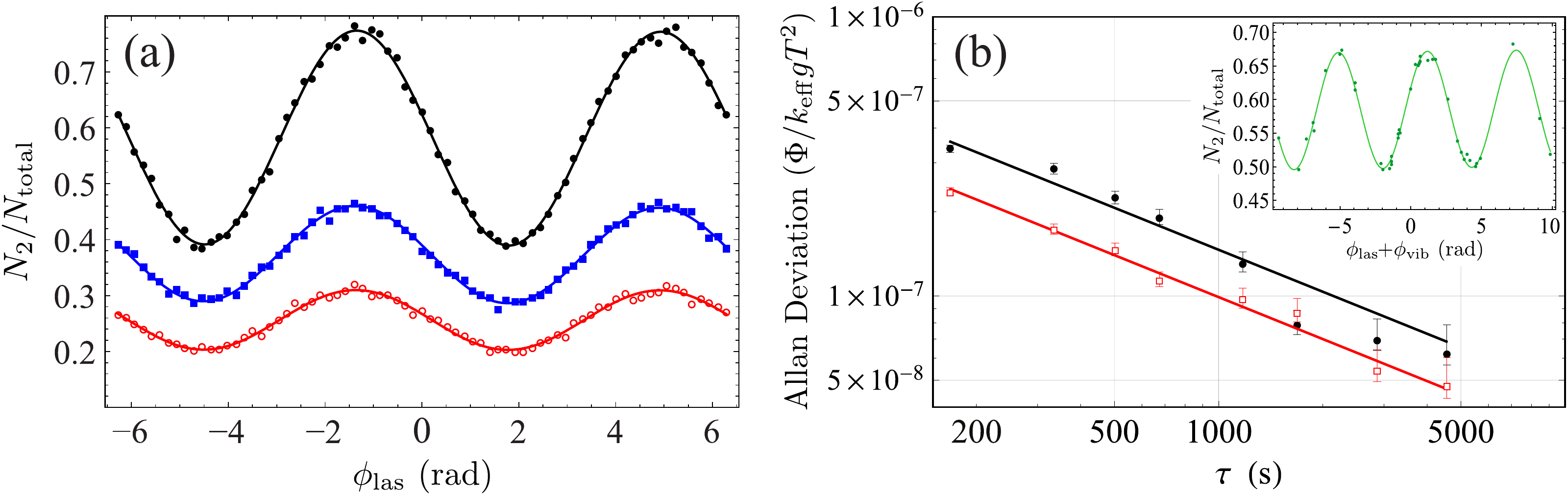}
  \caption{(a) Interference fringes for an interrogation time of $T = 100$ $\mu$s and a $\pi$-pulse duration $\tau_{\pi} = 6$ $\mu$s. Here, each point represents a single repetition of the interferometer as a function of a controlled phase shift $\phi_{\rm las}$ applied to the Raman laser before the final $\pi/2$-pulse. Atoms cooled to $\sim 16$ $\mu$K using a standard red molasses (red open circles) yield a fringe contrast of 10\% and SNR $\simeq$ 20. Cooling the atoms to $\sim 5$ $\mu$K in a gray molasses (blue filled squares) gives 17\% contrast and SNR $\simeq$ 30. Finally, adding a state preparation sequence after the gray molasses (black filled circles) yields a fringe contrast of 38\% and SNR $\simeq$ 45. The central fringe at $\phi_{\rm las} \simeq 1.7$ rad is phase-shifted due to gravity. (b) Allan deviation of phase measurements from the $^{39}$K gravimeter at $T = 22$ ms. Each measurement is composed of two interferometer fringes, one along each excitation direction, with each fringe containing 40 points (see inset). At a cycle time of $T_{\rm cyc} \simeq 2$ s, each measurement required $\tau = 160$ s---corresponding to the first point on the Allan deviation. Black filled circles correspond to the inertially-sensitive half-difference phase $\Delta\Phi$, while the half-sum phase $\Sigma\Phi$ is shown as red open squares, both of which are scaled by $k_{\rm eff} g T^2$ to obtain relative units of $g$. The bias drift of the mechanical accelerometer is corrected on the half-difference phase using our rubidium reference signal. Inset: typical interference fringe at $T = 22$ ms using upward excitation.}
  \label{fig:KGravimeter}
\end{figure*}

We characterize the quality of our state preparation sequence in terms of the purity and the temperature of the atoms in their target state. Figure \ref{fig:KSampleProperties}(a) shows a measurement of the relative population in the $\ket{F = 1}$ manifold obtained using velocity-insensitive Raman spectroscopy. Here, $N_2/N_{\rm total}$ is the fraction atoms in $\ket{F = 2}$ compared to the total atom number, and $\delta$ is the frequency difference between the Raman beams. After an initial depump pulse, the atoms are roughly equally distributed amongst the Zeeman states of $\ket{F = 1}$. The population in $\ket{F = 1, m_F = 0}$ increases from 39\% to 63\% after preparation with a single transfer sequence, and is further increased to 92\% by adding a push pulse, as shown in \Fig \ref{fig:KSampleProperties}(b). By cycling the transfer sequence the population in the target state can be increased, but saturates after 2-3 cycles around 75\%, as indicated by the black and green circles. This is consistent with our expectations based on a simple optical pumping model. By adding the push pulse we find further diminishing returns per cycle, and a saturation around 95\% after 2 cycles (red filled squares). We attribute this to additional pumping into $\ket{F = 1}$ during the push pulse. We also find that the temperature of the sample increases roughly linearly with each additional transfer sequence ($\sim 1$ $\mu$K per cycle) as a result of scattering-induced heating during the optical pumping pulses [see \Fig \ref{fig:KSampleProperties}(c)]. From these data there is clearly a trade off to be made between the purity of the sample and its temperature. Typically, we use two transfer sequences in our experiments in order to optimize the contrast of our atom interferometer signal.

\section{Applications in metrology}
\label{sec:Metrology}

We now discuss two ongoing precision measurements with cold potassium in the fields of inertial sensing and atomic spectroscopy. The first is an atom-interferometric gravimeter, which is part of a dual-species apparatus \cite{Barrett2015, Barrett2016, Lefevre2017} that aims to test Einstein's equivalence principle by comparing the gravitational acceleration of $^{87}$Rb and $^{39}$K. The second is a hyperfine spectroscopy measurement of the ground state splitting between the $4 S_{1/2}$ $\ket{F = 1}$ and $\ket{F = 2}$ ground states of $^{39}$K.

\subsection{Gravimetry}

We realize our potassium gravimeter using a Mach-Zehnder-like $\pi/2 - \pi - \pi/2$ sequence of Raman pulses \cite{Kasevich1991}, each separated by an interrogation time $T$ during which the atoms are in free fall. We demonstrate improvements made to the interferometer by measuring interference fringes with and without the gray-molasses cooling and state preparation steps, as shown in \Fig \ref{fig:KGravimeter}(a). These data were taken at a sufficiently small $T$ such that vibrations of the reference frame (defined by the mirror that retro-reflects the Raman beams) caused negligible phase noise. We observe a factor of $\sim 4$ increase in the contrast of the inertially-sensitive fringes by cooling the atoms in a gray molasses and preparing them in $\ket{F = 1, m_F = 0}$. The signal-to-noise ratio (SNR) also improves by more than a factor of 2, but is currently limited by technical noise from our detection system \cite{Antoni-Micollier2017}. Here, the gains in contrast and SNR are linked to the reduction of the velocity selection along the Raman beam's propagation direction. When increasing the interaction time, the transverse expansion of the atomic cloud contributes to the loss of contrast as the atoms experience a spatial variation of the Raman laser intensity \cite{Dickerson2013}. In this context, our preparation method is even more important in order to maintain a relatively high contrast and SNR at the timescales required for high-performance inertial sensors [see insert in \Fig \ref{fig:KGravimeter}(b)].

At larger interrogation times ($T \gtrsim 1$ ms), we apply a frequency chirp $\alpha$ to the Raman lasers in order to cancel the increasing Doppler shift of the two-photon resonance due to the atom's acceleration under gravity. The phase shift of the interferometer's central fringe is given by
\be
  \label{Phiupdown}
  \Phi_{\uparrow\downarrow} = \pm (\bkeff \cdot \bm{g} - \alpha) T^2 \pm \phi_{\rm vib} + \phi_{\rm sys}^{\uparrow\downarrow},
\ee
where $\bkeff \simeq (4\pi/\lambda) \hat{\bm{z}}$ is the effective Raman wavevector along the vertical direction, $\bm{g} = g\,\hat{\bm{z}}$ is the acceleration due to gravity, $\phi_{\rm vib}$ is a phase shift due to vibrations of the reference frame, and $\phi_{\rm sys}$ represents the total systematic shift. Since the counter-propagating Raman beams are generated by retro-reflecting a pair of co-propagating lasers, we can choose the direction of excitation (upward $\uparrow$, or downward $\downarrow$) via the Raman frequency during the first $\pi/2$ pulse. We reject systematic effects that are independent of this direction (\eg AC Stark shifts, quadratic Zeeman shifts) by alternating between upward and downward excitations for each repetition of the experiment \cite{Peters2001, Louchet-Chauvet2011a}. In the case where $\phi_{\rm sys}^{\uparrow\downarrow} \simeq \phi_{\rm sys}$ is independent of the excitation direction, the half-difference $\Delta\Phi \equiv \frac{1}{2}(\Phi_{\uparrow} - \Phi_{\downarrow}) = (\bkeff \cdot \bm{g} - \alpha) T^2 + \phi_{\rm vib}$ then yields a local measurement of $g$. Typically, one achieves this by suppressing phase noise due to vibrations, and precisely locating the central fringe for which $\alpha = \bkeff \cdot \bm{g}$. Similarly, the half-sum $\Sigma\Phi \equiv \frac{1}{2}(\Phi_{\uparrow} + \Phi_{\downarrow}) = \phi_{\rm sys}$ gives the total (direction-independent) systematic shift. Figure \ref{fig:KGravimeter}(b) displays the Allan deviation of gravity measurements obtained using a total interrogation time of $2T = 44$ ms \footnote{Our maximum interrogation time is limited by the geometry of our science chamber.}, where the inertially-sensitive half-difference phase ($\Delta\Phi$) is shown in black. These data show a sensitivity of approximately $2 \times 10^{-6}\, g/\sqrt{\rm Hz}$, and a statistical uncertainty of 60 $\mu$Gal after 80 minutes of integration.

A unique feature of our gravimeter is that we do not stabilize the reference mirror against vibrations, which causes a root-mean-squared phase noise of several radians at $T = 22$ ms. Instead, we measure these parasitic vibrations using a sensitive mechanical accelerometer (Nanometrics Titan) and later correct for the corresponding phase shifts on the interferometer \cite{LeGouet2008, Barrett2015, Fang2016}. An example of one such vibration-corrected interference fringe is shown in the inset of \Fig \ref{fig:KGravimeter}(b). We emphasize that without this correction, inertial phase measurements for $T$ greater than a few ms would not be possible with our apparatus. However, the mechanical accelerometer acts as a ``floating'' phase reference, since its output bias inevitably drifts due to temperature variations. This drift is then imprinted on our gravity measurements---frustrating long-term integration. This problem can be avoided by strongly filtering the DC component of the accelerometer signal, since only the AC component is used for correcting the vibration-induced phase shifts. For these data, we used the output of our simultaneous $^{87}$Rb gravimeter (sensitivity $\sim 30$ $\mu$Gal after 80 min.) as a phase reference for potassium in order to suppress the DC bias drift of the accelerometer \cite{Lefevre2017}. However, this method is also imperfect since any noise in the rubidium phase is imprinted on the potassium signal. We point out that this approach is equivalent to measuring the differential acceleration between the two species, and constitutes a test of the equivalence principle \cite{Schlippert2014a}.

An analysis of the half-sum phase gives an indication of the sensitivity level we could reach with a perfect phase reference, since this quantity is insensitive to inertial phases like $\phi_{\rm vib}$. We estimate this level to be $\sim 40$ $\mu$Gal after 80 minutes from the data shown in \Fig \ref{fig:KGravimeter}(b). Additionally, these data give a good indication of the stability of our experiment, since $\Sigma \Phi$ is a direct measure the relatively large one-photon light shifts and quadratic Zeeman shifts in potassium.

State-of-the-art gravimeters based on cold $^{87}$Rb sources can now achieve sub-$\mu$Gal uncertainty after only a few minutes of integration, with accuracies on the order of a few $\mu$Gal \cite{Gillot2014, Fang2016, Freier2016}. However, for potassium, the most precise gravity measurements demonstrated so far are on the order of 490 $\mu$Gal after one hour of averaging \cite{Schlippert2014a}. With our low-temperature, high-purity $^{39}$K sample, we have improved upon these results by about an order of magnitude. In terms of accuracy, we estimate an overall systematic uncertainty of $\sim 50$ $\mu$Gal, where the dominant contributions arise from the two-photon light shift \cite{Gauguet2009}, the curvature of the Raman beam wavefront \cite{Louchet-Chauvet2011a, Trimeche2017}, and the quadratic Zeeman effect \cite{Barrett2016}. Secondary contributions due to the one-photon light shift, gravity gradient and the Coriolis acceleration sum to less than 10 $\mu$Gal for $T = 22$ ms. Further details concerning systematic effects on our potassium gravimeter will be published elsewhere.

\subsection{Hyperfine spectroscopy}

\begin{figure}[!tb]
  \centering
  \includegraphics[width=0.48\textwidth]{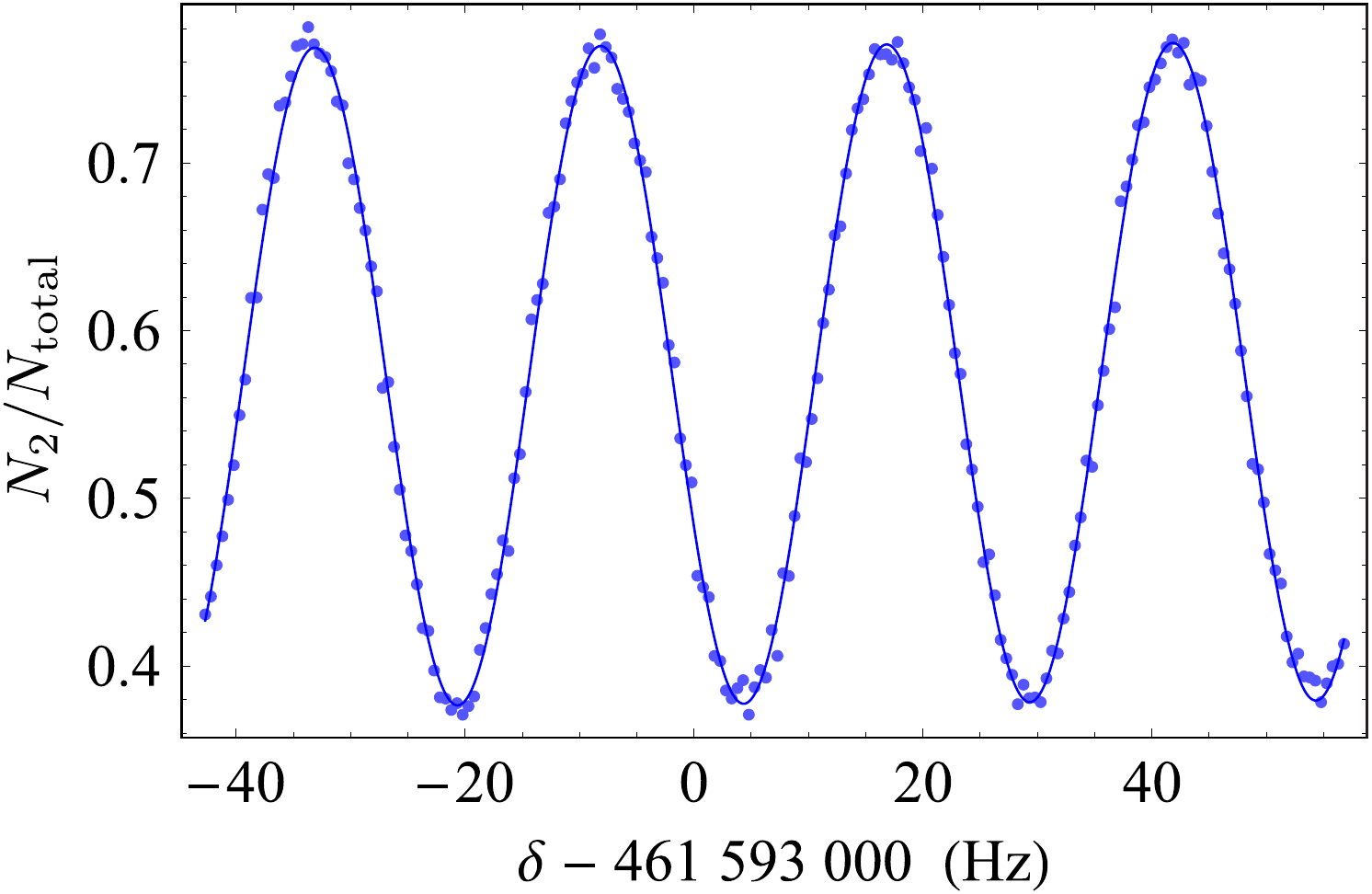}
  \caption{Optical Ramsey fringes using $T_R = 40$ ms and $\tau_R = 3$ $\mu$s. The corresponding sinusoidal fit (solid line) gives a relative statistical error of $\Delta f/f = 4.1 \times 10^{-11}$. Measurements were carried out near the peak of the Ramsey pattern where the SNR was largest. This peak is shifted by $-126.8(2)$ kHz from the central fringe corresponding to the hyperfine frequency splitting at 461.719720 MHz \cite{Arimondo1977} due to the AC Stark effect from the Raman beams. The corresponding shift of the central fringe is $11.2(1)$ Hz \cite{Zanon-Willette2016b}. Similarly, we estimate a shift due to the quadratic Zeeman effect of $-114.3(4)$ Hz.}
  \label{fig:KClock}
\end{figure}

We now describe an ongoing precision measurement of the clock transition in $^{39}$K using a Ramsey-type interferometer. The Ramsey technique \cite{Ramsey1949} involves a $\pi/2 - \pi/2$ pulse sequence, with pulse durations $\tau_R$ and free-evolution time $T_R$, during which the atoms are free-falling in a superposition of two internal states. As a function of the probe detuning $\delta$, the fraction of atoms in each state at the output of the spectrometer oscillates in the form of a Ramsey fringe pattern, where the fringe width scales inversely with $T_R$. The central fringe location---which remains fixed for all values of $T_R$ in the absence of systematic effects---gives a precise measure of the transition frequency between the two internal states. This type of interferometer is extensively used by the atomic clock community, either with microwave transitions in the case of alkali metals \cite{Kasevich1989, Gibble1993, Santarelli1999}, or with optical single-photon transitions in the case of alkaline and rare earth metals \cite{Nicholson2012, Hinkley2013}.

Instead of using a microwave field to probe $^{39}$K, our spectroscopy measurements were carried out using two pairs of co-propagating Raman beams---one pair travelling along the upward direction with $\sigma^+$ polarization and one downward with $\sigma^-$ polarization. In this configuration, both pairs of co-propagating Raman beams are velocity-insensitive and resonant with the $\ket{F=1,m_F=0} \to \ket{F=2,m_F=0}$ clock transition, while the counter-propagating pairs are velocity-sensitive and drive only $\Delta m_F = \pm 2$ transitions. These transitions are off-resonant due to both the Doppler and Zeeman effects. Figure \ref{fig:KClock} displays optical Ramsey fringes measured as a function of the frequency difference between Raman lasers. Using $T_R = 40$ ms, we obtain a statistical uncertainty in the fractional frequency of $\Delta f/f = 4.1 \times 10^{-11}$ with a single fringe measurement---more than a 30-fold improvement over the previous best measurements at $1.3 \times 10^{-9}$ \cite{Arimondo1977}. Although free from cavity-pulling effects \cite{Sortais2000}, our measurement is sensitive to light shifts from the Raman beams \cite{Zanon-Willette2016b}. In addition, we apply a relatively large bias field of $\sim 100$ mG in order to increase the efficiency of the state preparation scheme, but this too shifts the clock transition frequency due to the quadratic Zeeman effect. A preliminary study of the dominant systematic effects indicates our accuracy is limited at the level of a few $10^{-9}$ due to Zeeman shifts, AC Stark shifts, and the calibration of our frequency source \cite{Laurent1998, Stern2009}. We estimate frequency shifts due to cold collisions \cite{Kokkelmans1997, Sortais2000} to be less than 100 mHz at our densities of $\sim 5 \times 10^9$ atoms/cm$^3$.


\section{Conclusion}
\label{sec:Conclusion}

We have demonstrated that $^{39}$K can be used for high-precision measurements thanks to a new technique to prepare a cold sample of atoms in a magnetically-insensitive state with 95\% purity. We used this sample in an atom-interferometric gravimeter, and achieved a state-of-the-art performance. Potassium atoms prepared with this technique could be ideal for new determinations of the fine structure constant \cite{Bouchendira2011}, searches for new fundamental forces \cite{Jaffe2016}, or for tests of the equivalence principle \cite{Wolf2011a, Schlippert2014a, Lefevre2017}. We also demonstrate that, in a Ramsey-type spectrometer, the ground state hyperfine splitting of $^{39}$K can be measured with a sensitivity improved by more than a factor of 30. We anticipate that new rounds of measurements at increased interrogation time could yield a precision below $10^{-12}$. At this level, an improved mitigation strategy for the present systematic limitations will be required, including a characterization of collisional frequency shifts---which have not yet been measured in $^{39}$K. Our scheme can be easily modified to prepare atoms in any other ground state magnetic sub-levels, and is particularly well suited to atoms with closely-spaced excited states, such as the isotopes of lithium and potassium.

\acknowledgments

This work is supported by the French national agencies CNES (Centre National d'Etudes Spatiales), l'Agence Nationale pour la Recherche, the D\'{e}l\'egation G\'{e}n\'{e}rale de l'Armement, the European Space Agency, IFRAF (Institut Francilien de Recherche sur les Atomes Froids), and action sp\'{e}cifique GRAM (Gravitation, Relativit\'{e}, Astronomie et M\'{e}trologie). L. Antoni-Micollier  and B. Barrett thank CNES and IOGS for financial support. P. Bouyer thanks Conseil R\'{e}gional d'Aquitaine for the Excellence Chair. Finally, we wish to thank G. Santarelli for helpful discussions and M. Prevedelli for his assistance repairing a critical part of the experiment.

\bibliographystyle{apsrev4-1}
\bibliography{References}

\end{document}